\documentstyle[12pt,epsfig]{article}                                                             
                                                             
\newlength{\dinwidth}                                                             
\newlength{\dinmargin}                                                             
\def\lapproxeq{\lower .7ex\hbox{$\;\stackrel{\textstyle                                                             
<}{\sim}\;$}}                                                             
\def\gapproxeq{\lower .7ex\hbox{$\;\stackrel{\textstyle                                                             
>}{\sim}\;$}}                                                             
\def\be{\begin{equation}}                                                             
\def\ee{\end{equation}}                                                             
\def\bea{\begin{eqnarray}}                                                             
\def\eea{\end{eqnarray}}                                                             
                                                             
\def\funp{{I\!\!P}}                            
                                                            
\begin{document}                                                             
\titlepage                                                            
\begin{flushright}                                                             
IPPP/01/22 \\      
DCTP/01/44 \\                                                             
15 May 2001 \\                                                             
\end{flushright}                                                             
                                                            
\vspace*{2cm}                                                             
                                                            
\begin{center}                                                             
{\Large \bf Probabilities of rapidity gaps in high energy interactions} \\   
                                                            
\vspace*{1cm}                                                             
A.B. Kaidalov$^{a,b}$, V.A. Khoze$^a$, A.D. Martin$^a$ and M.G. Ryskin$^{a,c}$ \\                                                             
                                                           
\vspace*{0.5cm}                                                             
$^a$ Department of Physics and Institute for Particle Physics Phenomenology, University of      
Durham, Durham, DH1 3LE \\         
$^b$ The Institute for Theoretical and Experimental Physics (ITEP), 117259 Moscow, Russia   
\\                                                         
$^c$ Petersburg Nuclear Physics Institute, Gatchina, St.~Petersburg, 188300, Russia                    
\end{center}                                                            
                                                            
\vspace*{2cm}                                                             
                                                            
\begin{abstract}                                                             
We show that high energy hadronic reactions which contain a rapidity gap and a hard   
subprocess have a specific dependence on the kinematic variables, which results in a  
characteristic behaviour of the survival probability of the gap.  We incorporate this  
mechanism in a two-channel eikonal model to make an essentially parameter-free estimate of diffractive  
dijet production at the Tevatron, given the diffractive structure functions measured at HERA.   
The estimates are in surprising agreement with the measurements of the CDF collaboration.   
We briefly discuss the application of the model to other hard processes with rapidity gaps.  
\end{abstract}                                                  
             
\newpage                   
\section{Introduction}           
   
There has been much interest in the probability of rapidity gaps in high energy interactions to   
survive, since they may be populated by secondary particles generated by rescattering   
processes, see, for example, \cite{A1}--\cite{A8}.  The effect can be described in terms of  
screening or  
absorptive corrections.  To the best of our knowledge, the term survival probability was  
introduced by Bjorken \cite{A2} who estimated the probability using   
\be   
\label{eq:a1}   
S^2 \; = \; \frac{\int \: |{\cal M} (s, b) |^2 \: e^{- \Omega (b)} \: d^2 b}{\int \: | {\cal M} (s, b)   
|^2 \: d^2 b},   
\ee   
where ${\cal M}$ is the amplitude (in impact parameter $b$ space) of the particular   
process of interest at centre-of-mass energy $\sqrt{s}$.  $\Omega$ is the opacity (or optical   
density) of the interaction of the incoming hadrons\footnote{That is $i [1 - \exp (- \Omega/2)]$  
is the usual elastic scattering amplitude in impact parameter space.  $\Omega/2$ is frequently  
called the eikonal.}.   
   
It is perhaps more accurate to use the term \lq\lq suppression factor\rq\rq\ of a hard process  
accompanied by a rapidity gap, rather than \lq\lq survival probability\rq\rq.  It depends not  
only on the probability of the initial state to survive, but is sensitive to the spatial distribution  
of partons inside the incoming hadrons, and thus on the dynamics of the whole diffractive  
part of the scattering matrix.  It is important to  
note that the suppression factor $S^2$ is not universal, but depends on the particular hard  
subprocess, as well as the kinematical configurations.  In particular, $S^2$ depends on the  
nature of the colour-singlet (Pomeron or $W/Z$ boson or photon) exchange which generates  
the gap as well as on the distributions of partons inside the proton in impact parameter space  
\cite{KMR,KMR4,KMRH,GLM}.  In this paper we emphasize the importance of the dependence on the characteristic 
momentum fractions carried 
by the active partons in the colliding hadrons.  This leads to a much richer structure of the  
probability of rapidity gaps in processes mediated by colour-singlet $t$-channel exchange.   
The framework was introduced long ago\footnote{Reviews can be  
found, for example, in Refs.~\cite{B3,B4}.} \cite{B1,B2}, but only with the   
advent of rapidity gap events being observed in hard processes at the Tevatron and at HERA,   
is this rich physics now revealing itself.    
  
In Section~2 we briefly review the general framework, and, in particular, discuss a   
two-channel partonic model of diffraction.  Measurements of diffractive dijet production with a leading 
antiproton have been made recently by the CDF collaboration \cite{CDF} at the Tevatron.  This is an ideal  
process with which to compare the specific predictions of the models for high energy diffraction.  
In Section~3 we specify the partonic structure of the diffractive two-channel eigenstates.  
To set the scene for our main study we first, in Section~4, discuss diffractive dijet 
production assuming, for the moment, that rescattering  
corrections may be neglected.  As was emphasised in Ref.~\cite{CDF}, the calculation of the  
cross section, based on factorization in terms of diffractive structure functions obtained from  
HERA data, indicates a large discrepancy with the CDF measurements --- both in the  
normalisation and in the shape of the observed distribution.  The calculation lies about a  
factor of 10 above the data; the precise discrepancy depends on the kinematic domain.  In  
Section~5 we include rescattering corrections.  Clearly these will decrease the predictions,  
since now the rapidity gaps may be populated by secondary particles.  To allow for  
rescattering we use the two-channel eikonal model, reviewed in Sections~2 and 3, with parameters  
previously determined in a global description of the total, elastic and soft diffractive data  
available in the ISR to Tevatron energy range \cite{KMR}.  In this way we are able to make an  
essentially parameter-free prediction of both the normalisation and the shape of the CDF diffractive dijet  
data.  In Section~6 we discuss the application of the model to other hard processes with  
rapidity gaps, but on a less quantitative level than for dijet production.  In all cases 
the specific rescattering corrections are in the direction to improve the 
description of the data.  Finally, in Section~7, we present our conclusions.  
   
\section{Inelastic diffraction and diffractive eigenstates}   
   
In order to deduce the behaviour of inelastic diffraction, we start with the $s$-channel  
unitarity relation, which interrelates the proton-proton total cross section, elastic and inelastic  
scattering.  The unitarity relation is, in fact, valid at each value of the impact parameter  
separately, that is   
\be   
\label{eq:a2}   
2 {\rm Im} \: T_{fi} \; = \; \sum_n \: T_{nf}^* \: T_{ni},   
\ee   
where $T_{fi} (s, b)$ is the transition amplitude to go from state $i$ to state $f$.   
   
We follow a presentation by Pumplin \cite{B4}, after the original interpretation of Good and   
Walker \cite{B1}.  First we introduce states $\phi_k$ which diagonalize the diffractive part   
of the $T$ matrix.  Such eigenstates of diffraction only undergo elastic scattering.  Let us   
denote the orthogonal matrix which diagonalizes Im~$T$ by $C$, so that    
\be   
\label{eq:a3}   
{\rm Im}~T \; = \; CFC^T \quad\quad {\rm with} \quad\quad \langle \phi_k |F| \phi_j \rangle \;  
= \; F_j \:   
\delta_{jk}.   
\ee   
Now consider the diffractive dissociation of an arbitrary incoming state    
\be   
\label{eq:a4}   
| i \rangle \; = \; \sum_k \: C_{ik} \: | \phi_k \rangle.   
\ee   
The elastic scattering amplitude for this state satisfies   
\be   
\label{eq:a5}   
\langle i |{\rm Im}~T| i \rangle \; = \; \sum_k \: |C_{ik}|^2 \: F_k \; = \; \langle F \rangle,    
\ee   
where $F_k \equiv \langle \phi_k |F| \phi_k \rangle$ and where the brackets of $\langle F   
\rangle$ mean the average of $F$ over the initial probability distribution of diffractive   
eigenstates.  After the diffractive scattering described by $T_{fi}$, the final state $| f \rangle$   
will, in general, be a different superposition of eigenstates than those of 
$| i \rangle$ shown in (\ref{eq:a4}).  Suppose for simplicity, we neglect the real parts of the 
diffractive amplitudes, then  
\bea   
\label{eq:a6}   
\frac{d \sigma_{\rm tot}}{d^2 b} & = & 2 \: {\rm Im} \langle i |T| i \rangle \; = \; 2 \: \sum_k   
\: |C_{ik}|^2 \: F_k \; = \; 2 \langle F \rangle \nonumber \\   
& & \nonumber \\   
\frac{d \sigma_{\rm el}}{d^2 b} & = & \left | \langle i |T| i \rangle \right |^2 \; = \; \left (   
\sum_k \: |C_{ik}|^2 \: F_k \right )^2 \; = \; \langle F \rangle^2 \\    
& & \nonumber \\   
\frac{d \sigma_{\rm el \: + \: SD}}{d^2 b} & = & \sum_k \: \left | \langle \phi_k |T| i \rangle   
\right |^2 \; = \; \sum_k \: |C_{ik}|^2 \: F_k^2 \; = \; \langle F^2 \rangle. \nonumber   
\eea   
It follows that the cross section for the single diffractive dissociation of a proton,  
\be   
\label{eq:a7}   
\frac{d \sigma_{\rm SD}}{d^2 b} \; = \; \langle F^2 \rangle \: - \: \langle F \rangle^2,   
\ee   
is given by the statistical dispersion in the absorption probabilities of the diffractive   
eigenstates.     
   
Note that if all the components $\phi_k$ of the incoming diffractive state $| i \rangle$ were   
absorbed equally then the diffracted superposition would be proportional to the incident one   
and again the inelastic diffraction would be zero.  Thus if, at very high energies, the  
amplitudes $F_k$ at small impact parameters are equal to the black disk limit, $F_k = 1$,  
then diffractive production will be equal to zero in this impact parameter domain and so will  
only occur in the peripheral $b$ region.  This behaviour has already occurred in $pp$ (and  
$p\bar{p}$) interactions at Tevatron energies.  On the other hand, if there are, say, two  
diffractive channels with different eigenvalues, then the amount of inelastic diffraction  
increases with the spacing of the two eigenvalues.   
   
For instance, consider just two diffractive channels \cite{BTM,GLM,KMR} (say, $p, N^*$),   
and assume, for simplicity, that the elastic scattering amplitudes for the two channels are   
equal.  Then the $T$ matrix has the form   
\be   
\label{eq:a8}   
{\rm Im}~T \; = \; 1 \: - \: e^{- \Omega/2},  
\ee   
where the eikonal matrix $\Omega$ has elements   
\be  
\label{eq:a9}   
\Omega_{f^\prime i^\prime}^{fi} \; = \; \Omega_0 \: \omega^{fi} \: \omega_{f^\prime   
i^\prime}.  
\ee  
The individual $\omega$ matrices, which correspond to transitions from the two incoming   
hadrons, each have the form  
\be  
\label{eq:b9}  
\omega \; = \; \left ( \begin{array}{cc} 1 & \gamma \\ \gamma & 1 \end{array} \right ).  
\ee  
The parameter $\gamma (s, b)$ determines the ratio of the inelastic to elastic transitions.  The   
overall coupling $\Omega_0$ is also a function of the energy $\sqrt{s}$ and the impact   
parameter $b$.  
  
With the above form of $\omega$, the diffractive eigenstates are  
\be  
\label{eq:c9}  
| \phi_1 \rangle \; = \; \frac{1}{\sqrt{2}} \left (| p \rangle + | N^* \rangle \right ), \quad\quad |   
\phi_2 \rangle \; = \; \frac{1}{\sqrt{2}} \left (| p \rangle - | N^* \rangle \right ).  
\ee  
In this basis, the eikonal has the diagonal form  
\be  
\label{eq:d9}  
\Omega_{m^\prime n^\prime}^{mn} \; = \; \Omega_0 \: r^{mn} \: r_{m^\prime n^\prime},  
\ee  
where $m, n = \phi_1, \phi_2$ and  
\be  
\label{eq:e9}  
r \; = \; \left ( \begin{array}{cc} 1 + \gamma & 0 \\ 0 & 1 - \gamma \end{array} \right ).  
\ee  
In the case where $\gamma$ is close to unity, $\gamma = 1 - \varepsilon$, one of the   
eigenvalues is small.  

\section{Parton configurations of the diffractive eigenstates}

The simple two-channel model of Section~2 allows the prominent features of hard diffractive 
processes to be explained, which are beyond the scope of the single channel eikonal.  First we 
note that the parameter $\gamma$, which determines the ratio of inelastic to elastic transitions, 
needs to be in the range 0.4--0.6 to be in accord with the experimental data on diffractive 
dissociation at moderate energies.  Thus we know that there will be a big difference ($1 \pm \gamma$) 
in the absorptive cross sections for scattering in the two diffractive eigenstates.  To be specific, 
in this work we use the results of the detailed analysis of the elastic and soft diffractive data 
that was presented in Refs.~\cite{KMR,KMR4}.  There $\gamma$ was taken to be 0.4.
   
In QCD the diagonal states correspond to quark and gluon configurations with different  
transverse coordinates\footnote{Partonic models of diffraction were originally introduced in  
Refs.~\cite{VHF,PU}.}.  For small transverse size $r$ such (colourless) configurations  
interact as small colour dipoles with total interaction cross sections $\sim r^2$.  Thus, to a  
rough approximation, we can separate all the parton configurations of the colliding hadrons  
into those with small size and those with large size.  In our two-channel  
example above these would correspond to the states $| \phi_2 \rangle$ and $| \phi_1 \rangle$  
respectively.   
   
It is informative to discuss the phenomenon in terms of the usual Reggeon   
diagrams.  Assume that some \lq\lq hard\rq\rq\ diffractively produced state\footnote{The state   
$h$ should really be regarded as a third diffractive channel, but such a new state with a small  
production cross section gives a negligible contribution to $\Omega$.} \lq\lq $h$\rq\rq\ is  
strongly coupled to state $| \phi_2 \rangle$ and weakly   
to $| \phi_1 \rangle$.  It follows from (\ref{eq:c9}) that the Pomeron couplings of $h$ to $p$   
and $N^*$ satisfy   
\be   
\label{eq:a10}   
g_{ph}^\funp \; = \; -g_{N^* h}^\funp,   
\ee   
and that the $p$ and $N^*$ intermediate states for double-Pomeron exchange contribution   
(Fig.~1) interfere destructively, since   
\be   
\label{eq:a11}   
g_{pp}^\funp \: g_{ph}^\funp \: + \: g_{pN^*}^\funp \: g_{N^* h}^\funp \; = \; g_{pp}^\funp   
\: g_{ph}^\funp (1 - \gamma).  
\ee   
The cancellation which occurs for $\gamma \approx 1$, happens, in this simple model, for all   
multiple-Pomeron exchanges.  This phenomenon of \lq\lq colour transparency\rq\rq\ for   
small-size configurations has been known for a long time \cite{CT}.  

\begin{figure}[htb]  
\begin{center}
\
\epsfig{figure=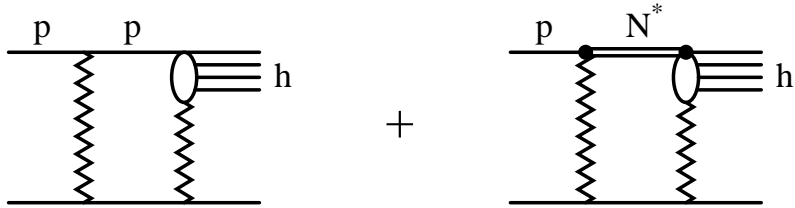,height=1.1in}  
\caption{The double-Pomeron exchange contribution to diffractive $h$ production 
in the simple two-channel model of (\ref{eq:e9}).
\label{fig:fig1}}  
\end{center}
\end{figure}
   
In order to specify the diffractive eigenstates $|\phi_1 \rangle$ and $|\phi_2 \rangle$ we shall 
consider two simple models (A and B).  It is natural to identify 
the component with the smaller absorptive cross section (that is $|\phi_2 \rangle$) with 
the state which contains less partons and which has a large typical momentum fraction $x$ for each 
parton.  From the QCD viewpoint, the small size component of the proton (where all the valence 
quarks are close together) has the smallest absorptive cross section, due to colour transparency.  
From the Regge viewpoint, the component with the largest absorptive cross section corresponds to 
the eigenstate ($|\phi_1 \rangle$) with a larger number of partons in the small $x$ region.  Thus 
from both viewpoints we expect the component with the smaller cross section (smaller transverse 
size) to have a larger average $x$ of each parton.  At the moment, it is impossible to be more 
specific, and so to make numerical estimates we consider two alternatives.

First, in model A, we identify the valence quarks with $|\phi_2 \rangle$ with the smaller absorption, 
and the gluons and sea quarks with $|\phi_1 \rangle$.  Of course the model is oversimplified.  It is clear 
that there is a part of the valence component with large size, while on the other hand the gluons 
and sea quarks contribute to the small size component.  In general, one can write each partonic 
distribution $f_i (x, Q^2)$ ($i$ = valence, sea, glue) as the sum of a small ($S$) and large 
($L$) size component
\be
\label{eq:b11}
f_i (x, Q^2) \; = \; f_i^S (x, Q^2) \: + \: f_i^L (x, Q^2).
\ee
In a model, where the probabilities of the $S$ and $L$ components in the proton are equal,  
as in Section~2, these components should satisfy the following sum rules,
\bea
\label{eq:c11}
\int_0^1 \: dx \: f_V^S (x, Q^2) & = & \int_0^1 \: dx \: f_V^L (x, Q^2) \; = \; \frac{3}{2} \\
& & \nonumber \\
\label{eq:d11}
\int_0^1 \: dx \: x \sum_i \: f_i^S (x, Q^2) & = & \int_0^1 \: dx \: x \sum_i \: f_i^L 
(x, Q^2) \; = \; \frac{1}{2},
\eea
which follow from the conservation of valence quark number and energy respectively.

We can therefore introduce an alternative model in terms of modified parton distributions
\be
\label{eq:e11}
f_i^{S,L} (x, Q^2) \; = \; P_i^{S,L} (x, Q^2) \: f_i (x, Q^2),
\ee
where the projection operators have the simple forms
\be
\label{eq:f11}
P_i^L \; = \; (1 - x)^{n_i (Q^2)}, \quad\quad P_i^S \; = \; 1 \: - \: P_i^L.
\ee
We determine the values of $n_i$ in order to satisfy the sum rules of (\ref{eq:c11}) and 
(\ref{eq:d11}).  We call this model~B.  It turns out that both models A and B give rather similar 
predictions.  We study the implications of the models in Section~5.

\section{Diffractive dijet production --- a first look}    
  
Recently CDF have measured diffractive dijet production for events with a leading antiproton 
at the Tevatron \cite{CDF}.  These observations, coupled with the diffractive measurements by H1 
\cite{H1} and ZEUS \cite{ZEUS} at HERA, offer the opportunity to explore the diffractive framework in some   
detail.  The processes are shown schematically in Fig.~2, in the absence of rescattering   
corrections.  The lower parts of the diagrams, shown as Pomeron exchange, are to be   
understood as including multiple Pomeron contributions.  

\begin{figure}[htb]  
\begin{center}
\
\epsfig{figure=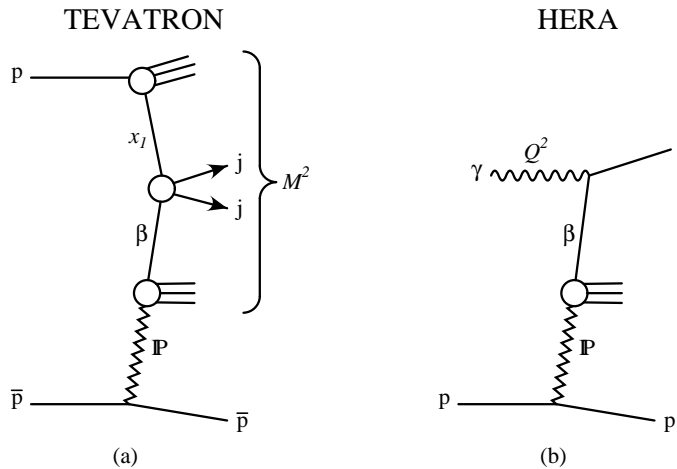,height=2.5in}  
\caption{Schematic diagrams for diffractive dijet production at the Tevatron and for   
diffractive deep inelastic scattering at HERA.  The rescattering corrections are omitted in   
these diagrams.
\label{fig:fig2}}  
\end{center}
\end{figure}
  
If we ignore rescattering corrections, for the moment, then the cross section for diffractive   
dijet production of Fig.~2(a), integrated over $t$, may be written as  
\be  
\label{eq:a12}  
\sigma \; = \; \sum_{i,k} \: \int \: F_\funp (\xi) \: f_i^\funp (\beta, E_T^2) \: 
f_k^p (x_1, E_T^2) \: \hat{\sigma} \: d \beta dx_1 d\xi,   
\ee  
where $\hat{\sigma}$ is the cross section to produce dijets from partons carrying longitudinal   
momentum fractions $x_1$ and $\beta$ of the proton and Pomeron respectively.  This would 
correspond to the Ingelman-Schlein conjecture \cite{IS}.  Information   
on the diffractive structure functions $f_i^\funp (\beta, Q^2)$ is obtained from measurements   
of the process of Fig.~2(b) at HERA \cite{H1,ZEUS}.  $F_\funp (\xi)$ is the flux factor taken   
for the Pomeron  
\be  
\label{eq:a13}  
F_\funp (\xi) \; = \; \int \: dt \: \frac{C_\funp e^{B t}}{\xi^{2 \alpha_\funp (t) - 1}},  
\ee  
where $\xi$ is the fractional momentum loss of the recoil antiproton.  In Regge theory, the   
coupling satisfies $C_\funp = (g^\funp_{pp})^2/16 \pi$, such that the total $p\bar{p}$ cross   
section is given by  
\be  
\label{eq:a14}  
\sigma_{\rm tot} (p\bar{p}) \; = \; (g_{pp}^\funp)^2 \: (s/s_0)^\Delta,  
\ee  
where $s_0 \equiv 1~{\rm GeV}^2$ and $\Delta = \alpha_\funp (0) - 1$.  When 
$\xi$ is not too small the contribution of secondary Reggeons must be added.  
  
CDF present measurements of the ratio of dijet production for $E_T ({\rm jet} 1, {\rm jet} 2) > 7~{\rm GeV}$ 
with a rapidity gap to that without   
a gap as a function of $x = \beta \xi$ (the fractional longitudinal momentum of the $\bar{p}$   
carried by the parton), for six $\xi$ bins in the range $0.035 < \xi < 0.095$ with $|t| < 1~{\rm GeV}^2$ 
\cite{CDF}.  In the ratio, the terms $f_j^p (x_1, E_T^2) \hat{\sigma}$ cancel, assuming that single 
gluon $t$-channel exchange dominates the hard subprocess.  Hence the data determine the diffractive 
structure function of the antiproton\footnote{Here we define the 
Pomeron flux slightly differently to Ref.~\cite{CDF} by including $C_\funp$ in (\ref{eq:a13}).}  
\bea  
\label{eq:a15}  
\tilde{F}_{jj}^D & = & \frac{1}{\xi_{\rm max} - \xi_{\rm min}} \: \int_{\xi_{\rm   
min}}^{\xi_{\rm max}} \: d \xi \: F_\funp (\xi) \: \beta \left [f_g^\funp (\beta, E_T^2) \: + \:   
\textstyle{\frac{4}{9}} \: f_q^\funp (\beta, E_T^2) \right ] \nonumber \\  
& & \\  
& & \quad\quad\quad\quad\quad\quad\quad + \; {\rm secondary~Reggeon~contributions}.   
\nonumber  
\eea  
The CDF measurements of $\tilde{F}_{jj}^D$ are shown by the data points in Fig.~3,   
together with five curves representing predictions of $\tilde{F}_{jj}^D$ based on various sets of   
diffractive structure functions, themselves obtained by fitting to HERA diffractive data.  The   
structure functions $f_i^\funp (\beta, Q^2)$ are evaluated at $Q^2 = 75$~GeV$^2$, which   
approximately corresponds to the average $E_T^2$ of the CDF data.  

The prediction labelled by H1 is obtained from the H1 diffractive data, and corresponds to fit~2 
of the H1 collaboration \cite{H1}.  The curve labelled by ZEUS(Pom) corresponds to the 
prediction obtained from ZEUS data in Ref.~\cite{ZEUS}\footnote{Note that the 
curve in Fig.~3 for the ZEUS structure function differs from that calculated in \cite{RSBJP}.}.  
It does not include the contribution of secondary Reggeons.  Note that the ZEUS data are in the 
region of very small $\xi$ and thus are practically insensitive to these contributions.  The 
curve labelled ZEUS$^\prime$ includes the secondary Reggeon contribution as determined by H1 
collaboration\footnote{This procedure may not be completely consistent as values of the 
Pomeron intercept are different in the analyses of the H1 and ZEUS data (see Refs.~\cite{H1,ZEUS}).  
This can lead to a modification of the secondary Reggeon contribution for the ZEUS parametrization.}.  
A comparison of the two latter curves shows that the non-Pomeron \lq\lq background\rq\rq\ is rather 
important in the $\xi$ region covered by CDF (about 50\% of the total contribution).  These three 
predictions are representative of those obtained from the various sets of   
diffractive structure functions that are available \cite{REST}.  They illustrate the large   
uncertainties in the predictions of the shape of $\tilde{F}_{jj}^D$ at {\it large} $\beta$, and   
in the overall normalisation.  On the other hand, the shape predicted for $\beta \lapproxeq  
0.15$ is well determined to be $\beta^{-\delta}$ with $\delta = 0.4-0.5$, and differs markedly  
from the measured $\delta \simeq 1$ behaviour of the CDF data.  
  
\begin{figure}[htb]  
\begin{center}
\
\epsfig{figure=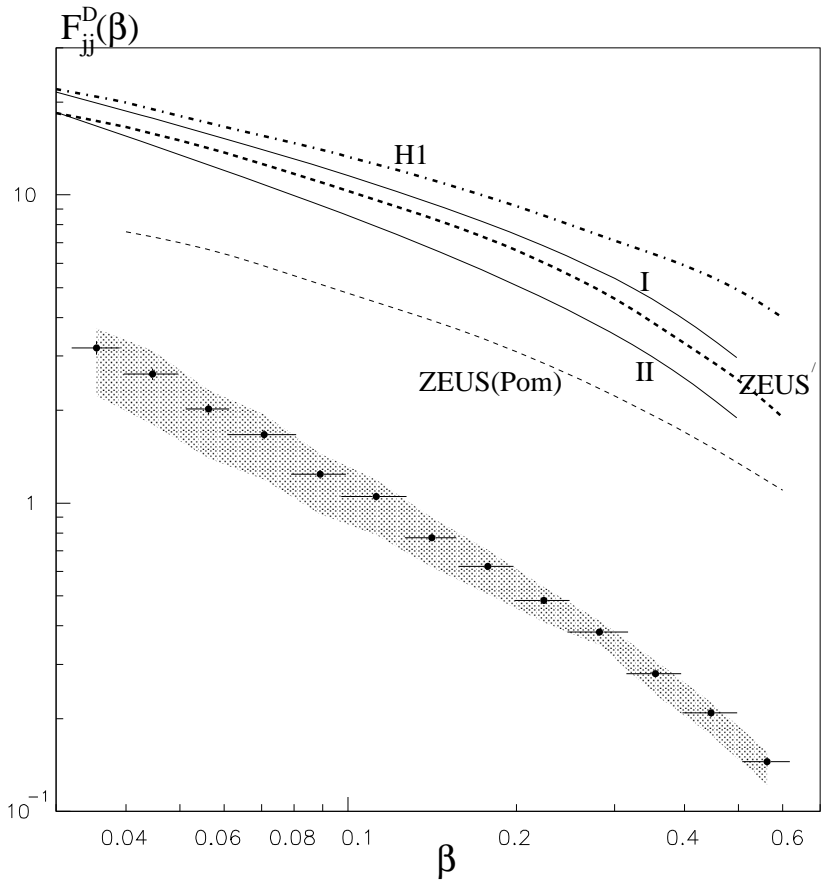,height=5in}  
\caption{A comparison of the measured CDF dijet diffractive distribution as a function  
of $\beta$, with different predictions obtained from analyses of HERA diffractive data 
assuming Regge factorization and that rescattering corrections are neglected.  
The shaded region on the CDF data shows the band of uncertainty shown in Ref.~\cite{CDF}.
\label{fig:fig3}}  
\end{center}
\end{figure}

The diffractive gluon distribution is the main contributor to the predictions of   
$\tilde{F}_{jj}^D$.  Although the diffractive quark distributions are well-measured at   
HERA (since the photon couples directly to the quark), the gluon distribution is determined   
from the detailed $Q^2$ behaviour of the experimental data using QCD evolution.   
Moreover, the uncertainties in the diffractive structure functions $f_i^\funp (\beta, Q^2)$ are  
amplified by sizeable differences   
between H1 and ZEUS diffractive data in certain $\beta, Q^2$ domains.  These uncertainties   
mainly affect the gluon distribution and are responsible for the ambiguity in the predictions   
for the shape of $\tilde{F}_{jj}^D$ at large $\beta$, and in the overall normalisation.  All the   
predictions give similar shapes for $\beta \lapproxeq 0.15$, because they are determined mainly   
by the QCD evolution.  We will therefore study this evident difference between the CDF and   
HERA shape of $\tilde{F}^D$ at small $\beta$, as well as the difference in overall   
normalisation\footnote{Note that earlier CDF results \cite{DINO} on diffractive $W$ boson, dijet, $b$-quark 
and $J/\psi$ production rates, using forward rapidity gap tagging, have already provided evidence against 
approaches which do not account for rescattering effects.}, which are clearly not reproduced in the 
naive model based on Fig.~2.  

Although we see from the curves shown in Fig.~3 that, at present, there are large uncertainties 
in the Pomeron structure function measured at HERA, curves I and II should provide a realistic 
illustration of the range of acceptable values, for the reasons given above.  These two alternative 
curves are used in the analysis of the CDF data presented below.  The curve I corresponds to 
the function $4.6 (1-\beta)^{1.1} \beta^{-0.45}$, and is very close to the parametrization of 
Capella et al.~\cite{KAID} (with secondary reggeons taken from \cite{H1}), while the curve II corresponds 
to the parametrization $2.5 (1-\beta) \beta^{-0.58}$, which we choose to account for 
possible variations due to the uncertainty in the secondary Reggeon contribution.  Recall both 
parametrizations corresponding to $Q^2 = 75~{\rm GeV}^2$.
  
The discrepancies between the Tevatron and HERA data were discussed in \cite{COMP}, where it 
was emphasized that the   
survival probability of the gap is (i) small, and (ii) dependent on the value of $\beta$.   
Physical arguments were presented which qualitatively reproduce the scale of normalisation  
and some trends in the $\beta$ dependence at large $\beta$.  Note that the effects causing the 
observed $\beta$-dependence of the diffractive structure function considered in \cite{COMP} and 
in this paper concern different regions in $\beta$ and are of different dynamical origin.  While 
the fall-off at $\beta \rightarrow 1$ is attributed in \cite{COMP} dominantly to Sudakov 
suppression effects, in this paper the variation of the shape of the $\beta$-distribution is 
explained mainly by the competition between the different parton configurations.  In this way, 
we present below a two-channel model prediction, based on \cite{KMR}, which turns out to be in 
surprising agreement with both the normalisation difference and in the shape of the distributions 
at low $\beta$.  
  
\section{Diffractive dijet production including rescattering \\ effects}   
  
To explain the main features of the CDF diffractive dijet data it is sufficient to consider the   
two-component diffractive models introduced in Section~3.  In model~A we assume that the sea   
quarks and gluons mainly occur in large-size configurations of the incident proton, while the   
valence quarks occupy predominantly small-size configurations.  This is, of course, an   
oversimplification of the real situation, but we find even this simple physical model is able to   
account for the behaviour of the data.  

The two-channel generalisation of (\ref{eq:a1}) gives, using model\footnote{In model B the subscripts 
\lq v\rq\ and \lq sea\rq\ correspond to the components with the smaller and larger absorption 
cross sections respectively.} A of Section~3, the survival probability of the   
gaps\footnote{In fact in the calculations a more accurate formula is used which takes into account the 
inelastic rescatterings of both of the colliding protons, see Appendix B of Ref.~\cite{KMR}.}  
\be  
\label{eq:a16}  
| S |^2 \; = \; \frac{\int \: d^2 b \left ( | {\cal M}_{\rm v}|^2 \: e^{- \Omega_{\rm v} (s,b)} \: +   
\: | {\cal M}_{\rm sea}|^2 \: e^{-\Omega_{\rm sea} (s,b)} \right )}{\int \: d^2 b \left ( |{\cal   
M}_{\rm v}|^2 \: + \: |{\cal M}_{\rm sea}|^2 \right )},  
\ee  
where ${\cal M}_{\rm v, sea}$ are the probability amplitudes (in impact parameter space) of   
the hard diffractive process corresponding to the valence quark and to the sea quarks and   
gluons respectively.  The functions $\Omega_i$ can be parametrized in the form\footnote{We   
show formula (\ref{eq:a17}) in order to again simplify the discussion.  In this 
simplified form the values of the parameters would be about $(g_{pp}^\funp)^2 = 25$~mb, $B_0 = 
8~{\rm GeV}^{-2}$, $\Delta = 0.1$ and $\alpha^\prime = 0.15~{\rm GeV}^{-2}$.  However, in 
practice we use the   
more realistic $\Omega_i (i = {\rm v, sea})$ that were determined in the global description of   
total, elastic and soft diffractive data in the ISR to Tevatron energy range \cite{KMR}.  In 
addition the pion-loop contribution in the Pomeron 
was included (that is the nearest $t$-channel singularity), as well as the contribution coming from   
large mass single- and double-diffractive dissociation.  These refinements are not  
crucial for the effects that we discuss here.}  
\be  
\label{eq:a17}  
\Omega_i \; = \; K_i \: \frac{(g_{pp}^\funp)^2 \: (s/s_0)^\Delta}{4 \pi B} \: e^{-b^2/4B},  
\ee  
with $i = {\rm v, sea}$, and where the slope of the Pomeron amplitude is  
\be  
\label{eq:a18}  
B \; = \; \textstyle{\frac{1}{2}} B_0 \: + \: \alpha^\prime \: \ln (s/s_0),  
\ee  
with $s_0 = 1~{\rm GeV}^2$.  We take $K_{\rm v} = 1 - \gamma$ and $K_{\rm sea} = 1 +   
\gamma$, consistent with the simple physical model introduced above.  The values of the other    
parameters were determined in a   
two-channel global description of the total, differential elastic and soft diffraction cross   
sections \cite{KMR}, in which the parameter $\gamma$ was fixed to be 0.4.  
  
First we indicate why the soft rescattering effects $(\Omega_i \neq 0)$ of the model based on   
(\ref{eq:a16}) modify the $\beta$ distribution of the dijet process in a characteristic way.    
Note that the CDF measurements cover a narrow $\xi$ interval, $0.035 \leq \xi \leq   
0.095$, and hence that the invariant mass squared of the diffractively produced state, $M^2 =   
\xi s$, remains close to the average value $2 \times 10^5~{\rm GeV}^2$.  Also the mass   
squared of the produced dijet system,  
\be  
\label{eq:a19}  
M_{jj}^2 \; = \; x_1 \beta M^2,  
\ee  
see Fig.~2, does not change much compared to its average value of about $1 \times 10^3~{\rm   
GeV}^2$ calculated for the CDF kinematical range.  Thus $x_1 \beta \simeq 0.005$ and so for 
$\beta \gapproxeq 0.25$ we have $x_1 \lapproxeq 0.02$, whereas for $\beta \sim 0.025$ we have $x_1 \sim 0.2$.  Therefore for large   
$\beta$ (small $x_1$) sea quarks and gluons will give the dominant contribution, while for   
small $\beta$ the valence quarks play an important role.  Hence the survival probability   
should increase as $x_1$ increases and $\beta$ decreases.  

\begin{figure}[!h]  
\begin{center}
\
\epsfig{figure=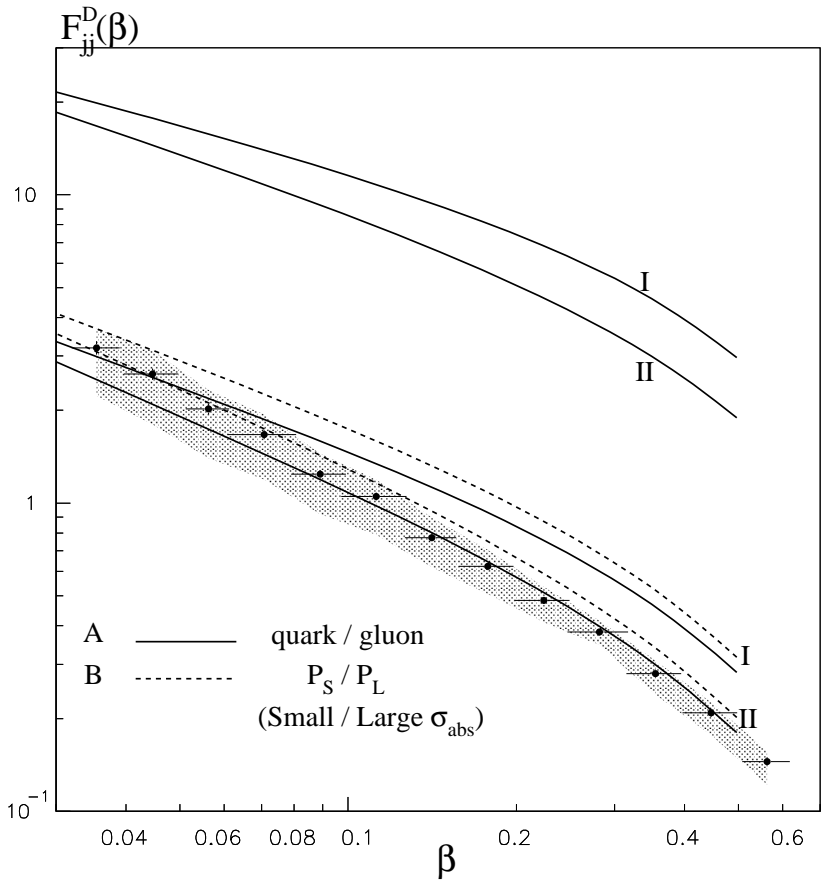,height=5in}  
\caption{The predictions for diffractive dijet production at the Tevatron, obtained from 
two alternative sets of \lq HERA\rq\ diffractive parton distributions I and II (of Fig.~3), 
compared with the CDF data \cite{CDF}.  The upper two curves correspond to the neglect of 
rescattering corrections, whereas the lower four curves show the effect of including these 
corrections using model~A (continuous curves) and model~B (dashed curves) for the diffractive 
eigenstates ($|\phi_i \rangle$ of Sections~2 and 3).}
\end{center}
\end{figure}
  
The calculation of the diffraction dijet rate, incorporating the rescattering effects of   
(\ref{eq:a16}), confirms these expectations, as shown by the lower pair of continuous curves 
(I and II) in Fig.~4.  These curves are parameter-free predictions of the diffractive dijet   
rate based on the two-channel eikonal model of Ref.~\cite{KMR} and on the diffractive 
distributions obtained from HERA data.  The two models (A and B of Section~3) for the 
diffractive eigenstates ($|\phi_1 \rangle$ and $|\phi_2 \rangle$) give similar predictions 
to each other, as shown respectively by the continuous and dashed curves in the lower part 
of Fig.~3.  We see that the pair of curves II satisfactorily reproduce the normalisation and 
the experimentally observed shape of the $\beta$ distribution.  Curves I also give a 
satisfactory description at low $\beta$; the difference at larger $\beta$ just reflects the 
uncertainty in the \lq HERA\rq\ diffractive distributions.  Recall that the predicted shapes 
show an anomalously strong increase\footnote{This increase is still somewhat weaker than that seen 
in the data $(\delta \approx 
1)$ \cite{CDF}.  However the CDF data include up to 4 jets, while the theoretical predictions are 
given for 2 jet production.  If the data are restricted to two jet production then the increase is 
less steep (and given by the lower part of the shaded band) \cite{CDF}, and, in fact, in agreement 
with our $\beta$ dependence.}, $1/\beta^\delta$ with $\delta \approx 0.8-0.9$, for small $\beta$, 
as compared with the $\delta \approx 0.4-0.5$ behaviour   
given by the partonic distributions of the Pomeron.  A possible change of $E_T$(jet), due to a 
variation $\Delta E_T$ of the transverse energy of the underlying event with $\beta$, was taken 
into account in our calculations.  We took $\Delta E_T = C (1 - \beta)^2$ with $C = 0.76$~GeV 
chosen so as to satisfy the observed $\langle \Delta E_T \rangle = 0.54$~GeV \cite{CDF}.  The 
origin of such $\beta$-behaviour can be traced to the fragmentation of the gluon jet.  It leads 
to a small $\sim 10\%$ decrease of the theoretical predictions for $\beta \gapproxeq 0.2$.  
  
The overall normalisation of the prediction for the CDF dijet data, which is reproduced by the   
average value of the survival probability (\ref{eq:a16}), is sensitive to the impact   
parameter distributions, ${\cal M}_i (s,b)$, of the hard diffractive process.  Such a comparison  
can therefore provide information on these distributions which, in turn, reveal the spatial  
structure of the hard process.  Our curves are obtained under the same assumptions for the  
single diffractive production of a massive hadronic state as were used in Ref.~\cite{KMR}.   
That is, as for the minimum bias single diffractive process, but without the term  
$\alpha^\prime \ln (M^2/s_0)$, since $\alpha^\prime \rightarrow 0$ in (LO) DGLAP  
evolution to the scale $\mu^2 \sim 75~{\rm GeV}^2$ of the hard subprocess.  

Our calculation of diffractive dijet production illustrates a crucial ingredient necessary in the   
description of rapidity gap processes.  Namely that the survival probability of a gap can   
depend on $x_1$ of the partons in the proton (see Fig.~2(a)).  This leads to many   
experimental consequences for processes with rapidity gaps.  For instance, if diffractive dijet   
production were measured at higher (LHC) energies with the same jet threshold $(E_T^j)$,   
then the values of $x_1$ of the partons from the proton will be much smaller throughout the   
same interval of $\beta$.  Thus the variation of $|S|^2$ with $\beta$ will disappear, and the   
shape of the $\beta$ distribution in this interval will be close to that measured at HERA.  The   
effect that is observed at the Tevatron is predicted to occur at the LHC, but at much smaller   
values of $\beta$, see (\ref{eq:a19}). 

\subsection{Other $\beta$ dependent effects}
\begin{figure}[htb]  
\begin{center}
\
\epsfig{figure=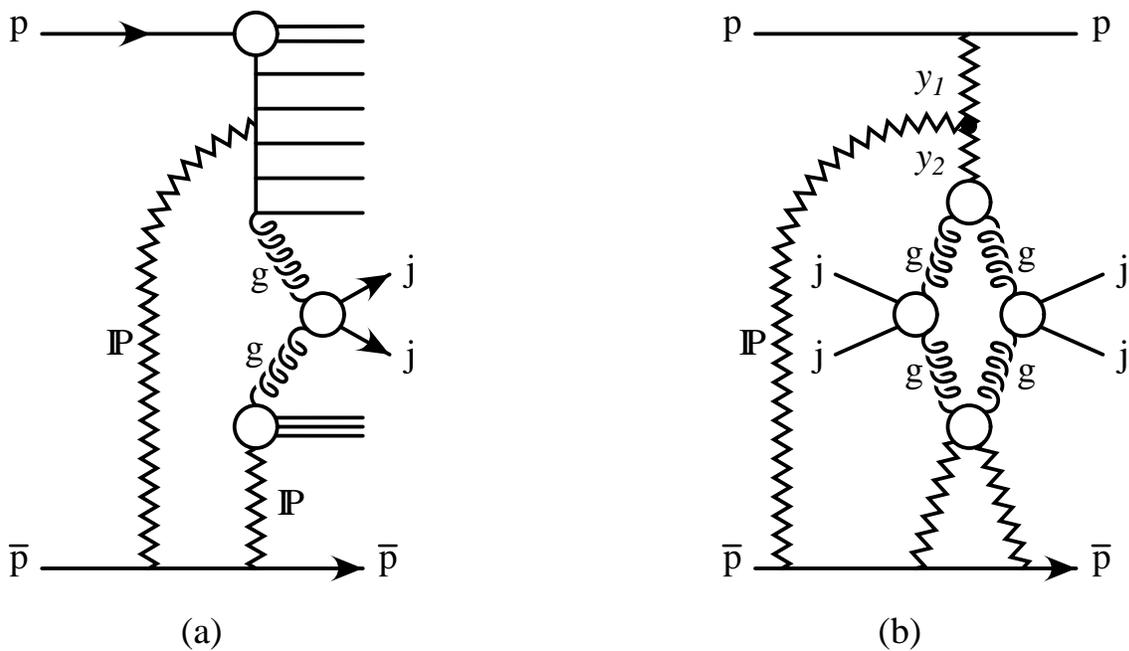,height=3.5in}  
\caption{
(a) Another possible contribution relevant to the  
calculation of the survival probability of the rapidity gap in diffractive dijet production, where  
the \lq rescattering\rq\ Pomeron couples to the upper \lq ladder\rq, rather than to the incoming  
proton. 
(b) The contribution to the diffractive dijet cross section coming from the interference  
of diagram (a) with the Born diagram of Fig.~2(a).\label{fig:fig4}}
\end{center}
\end{figure}

\vspace{-0.5cm}

We also estimated other possible mechanisms that may influence the predictions of the dijet 
$\beta$ distribution shown in Fig.~4.  We discuss these mechanisms in turn below.  None of them 
is expected to be significant, and anyway influences mainly the region of $\beta \sim 1$.  Most 
of the effects tend to make the $\beta$ distributions steeper and to improve the agreement with 
experiment.  However we have not included them in our predictions so as not to obscure the 
main phenomenon discussed in our paper.
\begin{itemize}
\item[(a)] The mechanism shown by the diagram of Fig.~5(a) describes the situation where the Pomeron 
couples to the \lq ladder\rq\ in the   
upper part of the diagram, rather than to the proton as we have considered so far.  It may 
influence the value of the survival probability at very small $x_1$.  The   
interference of Fig.~5(a) with the Born diagram of Fig.~2(a) leads to a contribution to the   
cross section shown in Fig.~5(b).  It contributes when the rapidity intervals $y_i$ (shown in   
Fig.~5(b)) are large; note that $y_1 + y_2 = \ln 1/x_1$.  This effect is small at the Tevatron   
because of the lack of phase space $(y_1 + y_2 \lapproxeq 5)$, but it should be taken into   
account at the LHC.  

\item[(b)] Another possible non-factorizable contribution is where a soft gluon from the Pomeron 
couples to the upper partons or spectator quarks of the proton \cite{CFS,COMP}.  Their dominant 
contribution may be summed and absorbed in the reggeization of the gluon, that is BFKL effects in 
parton evolution.  The remaining contribution is strongly suppressed since, when the soft $t$-channel 
gluon crosses an $s$-channel parton, it changes the colour structure of the corresponding splitting 
kernel.  For the singlet (gluon ladder) $N_C$ is replaced by $N_C/2$, whereas for the non-singlet 
(quark ladder) $C_F = (N_C^2 - 1)/2N_C$ is replaced by $-1/2N_C$.  Using the double $\log$ 
approximation we estimate this effect increases the prediction by less than 10\% for small $\beta$, 
and less than 4\% for $\beta > 0.3$.

\item[(c)] For large $\beta$ there is an additional Sudakov-like suppression due to QCD 
radiation from high $E_T$ jets \cite{COMP}.  Again the dominant contribution is already included 
in the effective structure of the Pomeron measured in DIS at HERA.  However conventional DGLAP 
evolution does not account for double $\log$s of the type $(\alpha_S/4\pi) \ln^2 (1 - \beta)$, 
which sum up to $\exp (- (\alpha_S/4\pi) C \ln^2 (1-\beta))$.  The effect may be important when 
$\beta \rightarrow 1$, but is negligible in the $\beta < 0.5$ domain of present interest.

\item[(d)] Hadronization may change the longitudinal momentum of the high $E_T$ jet \cite{COX}.  
The general effect is to shift the hadronic jet towards the centre-of-mass of the hadronic state, 
$M^2$ in Fig.~2(a), and to reduce the effective value of $\beta$.  A conservative estimate is that 
the prediction for $\tilde{F}_{jj}^D$ is changed by less than 10\% for $\beta < 0.3$, although it 
works in the desired direction to steepen the $\beta$ dependence.

\item[(e)] The predictions depend on the spatial size of the triple-Pomeron vertex.  The corresponding 
slope $b_p$ is small, but not well known.  We use the same slope $b_p = 1~{\rm GeV}^{-2}$ as in 
\cite{KMR}, but $b_p = 0$ or 2~GeV$^{-2}$ are not excluded.  Moreover we may expect $b_p$ to be 
smaller for larger $\beta \rightarrow 1$, when the Pomeron couples just to the hard sub-process.  
If we take $b_p = (1-\beta)~{\rm GeV}^{-2}$ then the prediction is unchanged in the small $\beta$ 
region, although it decreases by about 10\% at $\beta = 1/3$.  The sensitivity to the radius of the 
triple-Pomeron vertex indicates the importance of the experimental study of diffractive dissociation 
processes to better determine $b_p$.
\end{itemize}

\section{Predictions for other hard diffractive processes}  
  
The observation that the suppression factors can depend on the values of the momentum   
fractions $x_i$, carried by the partons in the colliding hadrons, has implications for hard   
diffractive-like processes in general.  For example, diffractive $W$-production at the  
Tevatron \cite{CDFW} is mediated dominantly by valence quarks in the proton, and  
hence the survival probability for such a process is comparatively large, $S^2 \simeq 0.2- 
0.3$.  On the other hand, for diffractive processes mediated by the gluonic components of the  
colliding hadrons (such as $b\bar{b}, J/\psi, \psi^\prime$ or $\Upsilon$ production) the  
survival probabilities should be smaller $S^2 \simeq 0.06-0.1$.  
  
An interesting application is to the production of two high $p_T$ jets $(\mbox{\boldmath $p$}_{1T} 
\simeq - \mbox{\boldmath $p$}_{2T})$ separated by a large rapidity gap, as   
measured by both the D0 \cite{D0} and CDF \cite{CDF2J} collaborations at the Tevatron at   
two energies, $\sqrt{s} = 630$ and 1800~GeV.  Both the quark and gluon components of the   
proton contribute in this case.  However, in our approach the suppression factor depends strongly 
on the type of parton (model~A) or on the $x$ value of the parton (model~B).  For a fixed energy 
$\sqrt{s}$, the ratio of the quark to the gluon component increases as $E_T$ of the jets increases, and   
as the rapidity interval $\Delta \eta$ between the jets increases.  The relative importance of   
the quark component also increases as the energy $\sqrt{s}$ decreases, simply due to   
kinematics.  These features of the simple   
two-channel model give effects which move in the right direction to explain outstanding   
puzzles in the interpretation of the D0 and CDF data for jets separated by a rapidity gap   
\cite{D0, CDF2J}.  In particular, they help to understand the $\sqrt{s}$, $E_T$ and $\Delta \eta$ 
dependences of the colour-singlet (rapidity gap) fraction measured at the Tevatron \cite{D0,CDF2J}.  
Our model gives a natural explanation of the observation by the D0 collaboration, that the suppression 
factor depends mainly on the $x$ of the partons and increases strongly with $x$, see 
Fig.~4(d) of \cite{D0}.  
  
All of the effects discussed above can be studied in hard diffractive processes in $p$-nucleus 
collisions at RHIC and LHC.  Investigation of the $A$-dependence can provide new information on 
the strength of shadowing effects.  Indeed for weak shadowing, the cross sections for {\it coherent} 
diffraction dissociation of a proton on nuclei behave as $\sim A^{4/3}$, while {\it incoherent} 
diffraction cross sections behave as $\sim A$.  In the opposite limit of very strong shadowing 
both cross sections have much weaker dependence on $A$, of the form $\sim A^{1/3}$.  Thus there is a 
strong change in $A$-dependence of diffractive production on nuclei depending on the strength of 
the shadowing effects. 
  
We also note that the survival probability for central Higgs production by $WW$ fusion, with   
large rapidity gaps on either side, is enhanced in the two-channel model in comparison with   
previous estimates \cite{KMR,KMRH}, which also included allowance for the survival  
probability.  Thus in Ref.~\cite{KMR} for $WW \rightarrow H$ process at the LHC $S^2$ was found 
to be 0.15, while the approach of this paper gives $S^2 = 0.24$.  This is an important process because 
it appears that large $q_T$ Higgs configurations can be chosen such as to identify the Higgs over the 
possible background processes at the LHC \cite{ZEP,KMR2}.  The same survival probability is applicable to central  
$Z$ boson production with a rapidity gap on either side, originating from $t$-channel gauge  
boson exchange.  Therefore $Z$ production at the LHC can be used to directly measure the  
survival probability of rapidity gaps relevant to Higgs production by $WW$ fusion \cite{A3}. 
  
\section{Conclusions}  
  
For hard processes with large rapidity gaps, we have demonstrated that the survival  
probability of the gaps has a much richer structure than is given by the simple one-channel  
eikonal approximation of (\ref{eq:a1}).  We introduced two-channel eikonal models in  
which {\it either} the valence quark and the sea quark (+ gluon) components of the proton have  
substantially different total cross sections of absorption $\sigma_i^{\rm tot} (s)$, which we 
called model~A, {\it or} alternatively, model~B, in which the small and large size diffractive 
components are specified according to sum rules (\ref{eq:c11}) and (\ref{eq:d11}).  The two models 
give similar results, and predict that the survival probability of the rapidity gap has a characteristic 
dependence on the kinematics of the process.  Data for diffractive dijet production at the 
Tevatron \cite{CDF}  enabled this kinematic dependence to be checked.  Taking the parameters of 
the two-channel models which were previously constrained in a global description \cite{KMR} of 
total, elastic and soft diffraction data, we calculated the $\beta$ distribution of diffractive 
dijet production.  The results are shown by the lower four curves in Fig.~4.  We see that there is 
general agreement between the predictions and the CDF measurements \cite{CDF}, both in normalisation 
and shape.  In fact, the use of Pomeron structure function II quantitatively reproduces the CDF 
data \cite{CDF}.  We emphasize that if the Pomeron structure function were known unambiguously then 
we would have an essentially unique prediction for the Tevatron data, demonstrated by the small 
difference between the predictions of models~A and B.  
  
Unfortunately, the agreement between the CDF diffractive dijet data and our calculations 
can be taken as a strong support for the low value of the survival probability $S^2$, which leads 
to the rather pessimistic expectations for the missing-mass Higgs search  
at the Tevatron \cite{A6,KMR01}.  

As precise data for other hard processes with rapidity gaps 
become available, it will be possible to refine the model and to identify the parton content of 
the diffractive eigenchannels.  We already showed that the simple two-channel model gave rescattering  
corrections which moved in the right direction to resolve discrepancies between the  
predictions and the data for processes which have so far been measured.  In this way, as  
precise data become available, it will be possible to perform a quantitative study to (i)  
determine the partonic content of the Pomeron, (ii) check the QCD evolution of the Pomeron 
structure functions, (iii) confirm the universality of the partonic decomposition, 
(iv) determine $\sigma_i^{\rm tot} (s)$ for the different diffractive eigenchannels, and 
(v) measure the impact parameter distributions of the \lq Born\rq\ amplitudes of the hard 
processes with rapidity gaps.  
  
\section*{Acknowledgements}  
  
We thank Dino Goulianos and Ken-ichi Hatakeyama for useful discussions.  One of us  
(VAK) thanks the Leverhulme Trust for a Fellowship.  ABK and MGR would like to thank the IPPP 
of the University of Durham for hospitality.  This work was partially supported by the UK 
Particle Physics and Astronomy Research Council, the EU Framework TMR programme, contract 
FMRX-CT98-0194 (DG 12-MIHT), NATO grant PSTCLG-977275 and also by grants RFBR~00-15-96610, 
00-15-96786, 01-02-17095 and 01-02-17383.


\newpage   
\begin{thebibliography}{xx}   
\bibitem{A1} Yu.L. Dokshitzer, V.A. Khoze and T. Sj\"{o}strand, Phys. Lett. {\bf B274}  
(1992) 116.  
\bibitem{A2} J.D. Bjorken, Int. J. Mod. Phys. {\bf A7} (1992) 4189; Phys. Rev. {\bf D47}  
(1993) 101.  
\bibitem{A3} H. Chehime and D. Zeppenfeld, Phys. Rev. {\bf D47} (1993) 3898.  
\bibitem{A4} R.S. Fletcher and T. Stelzer, Phys. Rev. {\bf D48} (1993) 5162.  
\bibitem{A5} E.M. Levin, {\tt hep-ph/9912402} and references therein.  
\bibitem{A6} V.A. Khoze, A.D. Martin and M.G. Ryskin, Eur. Phys. J. {\bf C19} (2001) 477 
and references therein.  
\bibitem{A7} M.M. Block and F. Halzen, Phys. Rev. {\bf D63} (2001) 114004.  
\bibitem{A8} M.G. Albrow and A. Rostovtsev, {\tt hep-ph/0009336} and references therein.  
\bibitem{KMR} V.A. Khoze, A.D. Martin and M.G. Ryskin, Eur. Phys. J. {\bf C18} (2000)   
167.  
\bibitem{KMR4} V.A. Khoze, A.D. Martin and M.G. Ryskin, Nucl. Phys. Proc. Suppl. {\bf 99} (2001) 525.
\bibitem{KMRH} V.A. Khoze, A.D. Martin and M.G. Ryskin, Eur. Phys. J. {\bf C14} (2000) 525.  
\bibitem{GLM} E. Gotsman, E. Levin and U. Maor, Phys. Lett. {\bf B452} (1999) 387; Phys.   
Rev. {\bf D60} (1999) 094011.  
\bibitem{B3} A.B. Kaidalov, Phys. Rep. {\bf 50} (1979) 157.  
\bibitem{B4} J. Pumplin, Physica Scripta {\bf 25} (1982) 191.   
\bibitem{B1} M.L. Good and W.D. Walker, Phys. Rev. {\bf 126} (1960) 1857.   
\bibitem{B2}V.N. Gribov, Sov. Phys. JETP {\bf 19} (1969) 483.  
\bibitem{CDF} CDF Collaboration:  T. Affolder et al., Phys. Rev. Lett. {\bf 84} (2000) 5043.  
\bibitem{BTM} K.G. Boreskov, A.M. Lapidus, S.T. Sukhorukov and K.A. Ter-Martirosyan,   
Nucl. Phys. {\bf B40} (1972) 397.  
\bibitem{VHF} L. Van Hove and K. Fialkowski, Nucl. Phys. {\bf B107} (1976) 211.  
\bibitem{PU} H.I. Miettinen and J. Pumplin, Phys. Rev. {\bf D18} (1978) 1696.  
\bibitem{CT} A.B. Zamolodchikov, B.Z. Kopeliovich and L.I. Lapidus, JETP Lett. {\bf 33}   
(1981) 595; \\  
G. Bertsch, S.J. Brodsky, A.S. Goldhaber and J.F. Gunion, Phys. Rev. Lett. {\bf 47} (1981)   
297.  
\bibitem{H1} H1 Collaboration:  T. Ahmed et al., Phys. Lett. {\bf B348} (1995) 681; \\  
C. Adloff et al., Z. Phys. {\bf C76} (1997) 613.  
\bibitem{ZEUS} ZEUS Collaboration:  M. Derrick et al., Z. Phys. {\bf C68} (1995) 569;   
Phys. Lett. {\bf B356} (1995) 129; Eur. Phys. J. {\bf C6} (1999) 43.  
\bibitem{IS} G. Ingelman and P. Schlein, Phys. Lett. {\bf B152} (1985) 256.
\bibitem{RSBJP} C. Royon, L. Schoeffel, J. Bartels, H. Jung and R. Peschanski,   
Phys. Rev. {\bf 63} (2001) 074004.  
\bibitem{REST} K.J. Golec-Biernat and J. Kwiecinski, Phys. Lett. {\bf B353} (1995) 329; \\  
T. Gehrmann and W.J. Stirling, Z. Phys. {\bf C70} (1995) 227.  
\bibitem{DINO} See K. Goulianos, Nucl. Phys. Proc. Suppl. {\bf 99} (2001) 37, for a recent review.
\bibitem{KAID} A. Capella, A. Kaidalov, C. Merino, D. Pertermann and J. Tran Thanh Van,   
Phys. Rev. {\bf D53} (1996) 2309.  
\bibitem{COMP} V.A. Khoze, A.D. Martin and M.G. Ryskin, Phys. Lett. {\bf B502} (2001) 87.  
\bibitem{CFS} J.C. Collins, L. Frankfurt and M. Strikman, Phys. Lett. {\bf B307} (1993) 161.
\bibitem{CDFW} CDF Collaboration:  F. Abe et al., Phys. Rev. Lett. {\bf 79} (1997) 2636.  
\bibitem{D0} D0 Collaboration: B. Abbott et al., Phys. Lett. {\bf B440} (1998) 189.  
\bibitem{CDF2J} CDF Collaboration:  F. Abe et al., Phys. Rev. Lett. {\bf 80} (1998) 1156;   
{\bf 81} (1998) 5278.  
\bibitem{COX} B. Cox, J.R. Forshaw and L. L\"{o}nnblad, {\tt hep-ph/9912489}.  
\bibitem{KMR2} V.A. Khoze, A.D. Martin and M.G. Ryskin, {\tt hep-ph/0104230}.  
\bibitem{ZEP} N. Kauer, T. Plehn, D. Rainwater and D. Zeppenfeld, Phys. Lett. {\bf B503} (2001) 113.
\bibitem{KMR01} V.A. Khoze, A.D. Martin and M.G. Ryskin, {\tt hep-ph/0103007}.
\end{thebibliography}
\end{document}